\documentclass[journal]{IEEEtran}
 \usepackage{amsmath,amstext,amsfonts,amssymb,mathrsfs}
\usepackage{graphicx,subfigure,algorithm,algorithmic,tikz}
\usepackage{hyperref}
\usepackage[sort,compress]{cite}
\usepackage{lipsum}

\newtheorem{theorem}{Theorem}

\newtheorem{definition}{Definition}

\makeatletter
\newcommand\fs@betterruled{%
  \def\@fs@cfont{\bfseries}\let\@fs@capt\floatc@ruled
  \def\@fs@pre{\vspace*{5pt}\hrule height.8pt depth0pt \kern2pt}%
  \def\@fs@post{\kern2pt\hrule\relax}%
  \def\@fs@mid{\kern2pt\hrule\kern2pt}%
  \let\@fs@iftopcapt\iftrue}
\floatstyle{betterruled}
\restylefloat{algorithm}
\makeatother

\usepackage{bbm}

\title{Online Algorithms for Dynamic Matching Markets in Power Distribution Systems}

\author{Deepan Muthirayan,~\IEEEmembership{Member,~IEEE}, Masood Parvania,~\IEEEmembership{Senior Member,~IEEE}, Pramod P. Khargonekar ~\IEEEmembership{Fellow,~IEEE}
\thanks{This work is supported in part by the National Science Foundation under Grant ECCS-1839429.
D. Muthirayan and P. P. Khargonekar are with the Department of Electrical Engineering and Computer Sciences, University of California Irvine, Irvine, CA (emails: deepan.m@uci.edu, pramod.khargonekar@uci.edu). M. Parvania is with the Department of Electrical and Computer Engineering, The University of Utah, Salt Lake City, UT 84112 (email: masood.parvania@utah.edu).}
}
\begin{document}

\maketitle
\thispagestyle{empty}
\pagestyle{empty}

\begin{abstract}
This paper proposes online algorithms for dynamic matching markets in power distribution systems. These algorithms address the problem of matching flexible loads with renewable generation, with the objective of maximizing social welfare of the exchange in the system.  
More specifically, two online matching algorithms are proposed for two generation-load scenarios: (i) when the mean of renewable generation is greater than the mean of the flexible load, and (ii) when the condition (i) is reversed. 
With the intuition that the performance of such algorithms degrades with increasing randomness of the supply and demand, two properties are proposed for assessing the performance of the algorithms. 
First property is convergence to optimality (CO) as the underlying randomness of renewable generation and customer loads goes to zero. The second property is
deviation from optimality, which is measured as a function of the standard deviation of the underlying randomness of renewable generation and customer loads. 
The algorithm proposed for the first scenario is shown to satisfy CO and a deviation from optimality that varies linearly with the variation in the standard deviation. 
We then show that the algorithm proposed for the second scenario satisfies CO and a deviation from optimality that varies linearly with the variation in standard deviation plus an offset under certain condition.
\end{abstract}


\section{Introduction}  \label{sec:int}
Electric power grids are undergoing a major transformation driven, to a significant extent, by the goal of decarbonization of energy systems, through large scale integration of renewable electricity sources (RES). 
RES are often some combination  of utility scale centralized and distributed wind and solar generators. 
Integration of RES in the operation and control of the grid is a significant challenge because photovoltaic (PV) solar and wind are highly uncertain, inherently variable, and largely uncontrollable. 
The information and decision complexity of managing very large numbers of the distributed resources is motivating research on decentralized control solutions \cite{annaswamy2013ieee}.

Market platforms in distribution systems facilitate decentralized management and control and can provide effective solutions for managing distributed RES. 
Essentially, such platforms can leverage the flexibility of loads to manage the variability of RES locally. This can allow the grid to be more locally self-sufficient and reduce the dependence on large centralized fossil fuel based generators. 
Management of such platforms poses interesting problems as customer loads and renewable generation are inherently random. Specifically, scheduling and matching of random supply and random demand is a critical and challenging problem \cite{parag2016electricity}. 

\subsection{Related Work}
There is a large body of work related to operation of distributed energy resources and flexible loads 
\cite{parisio2014model, shi2015real, khatami2018scheduling, khatami2018continuous, oikonomou2019deliverable, li2017real, sun2015distributed, strohle2016local, du2017distributed}. 
Authors in 
\cite{parisio2014model, shi2015real, khatami2018scheduling, khatami2018continuous, oikonomou2019deliverable} 
propose and study different algorithms for flexible loads and RES scheduling with the objective of minimizing the operation cost. 
The work in \cite{li2017real} provides asymptotic performance guarantees for an approach based on online stochastic optimization, while \cite{sun2015distributed} provides a theoretical analysis for a real-time algorithm for the objective of minimizing operational costs. 
The work in \cite{strohle2016local} develops a model for balancing flexible loads and local generation, and discuss its game theoretic properties but do not provide theoretical guarantees on its performance.
Authors in \cite{du2017distributed} propose a model predictive control scheme for minimizing customer dissatisfaction plus generation cost, but only provide experimental evaluation of their algorithm. 
In contrast to the above works, we develop online matching algorithms for matching flexible loads and local RES with the objective of maximizing social welfare. The key feature of our proposed model is the concept of criticality of flexible customers to capture their propensity to pay. In addition, our paper provides theoretical guarantees for the performance of the matching algorithms over a finite time horizon. 

Online matching has been extensively studied both in adversarial and stochastic settings in a variety of application domains \cite{karp1990optimal, kalyanasundaram1993online, mehta2005adwords, jaillet2013online} 
These papers provide algorithms that achieve a performance that is lower bounded. Algorithms with robust lower bounds in the online market clearing setting for a general commodity market without service constraints are provided in \cite{blum2006online}. 
In contrast to these works, our analysis captures the variation of the performance with respect to different scenarios of load and renewable generation, and allows us to better assess the performance of the algorithm across the scenarios.

\subsection{Contribution and Paper Structure}
The principal objective of this paper is to design online algorithms for dynamic matching of electricity energy in real-time operation of power distribution systems. The objective of the algorithms is to maximize social welfare (defined more precisely later).
Specifically, we provide a pair of online algorithms that are suitable for two distinct generation-consumption scenarios: (i) the mean of renewable generation is greater than the mean of the flexible loads, and (ii) when the inequality in (i) is reversed. 
The setting we study considers a novel flexible load model, which takes into account the deadline constraints and criticality of flexible loads in calculating the utility function of customers in the matching market. 
The criticality parameter of the proposed flexible load model specifies the rate at which a customer's willingness to pay for electricity decreases over time.
The proposed flexible load model is generic and suitable for modeling a range of flexible loads, such as electric vehicles (EVs), and flexible household appliances (e.g., dishwasher, washer, dryer).

A key question in designing online matching algorithms is: how should we assess the effectiveness of such algorithms across scenarios in managing random demand and random supply? 
In order to address this question, we propose a metric, termed competitive ratio (CR), to measure the relative performance of the proposed online algorithms. CR of an algorithm is the ratio of the expected social welfare of the algorithm and the expected social welfare of the oracle optimal algorithm. By definition, CR is less than or equal to 1. 
Intuitively, we expect that deviation of CR from 1 will be affected by the amount of randomness in renewable generation and load variability. 
With this intuition, we define the following concepts for assessing the proposed algorithms: (i) convergence to optimality, i.e., convergence of CR to one as the randomness decreases to zero, and (ii) deviation from optimality, where the deviation is measured as a function of the randomness of generation and loads (defined more precisely later).
Our online algorithm for the first scenario is shown to satisfy convergence to optimality and we also provide a lower bound for deviation from optimality. We show that this algorithm does not satisfy convergence to optimality for the second scenario. We then propose a modified algorithm for this case and provide results for its convergence and deviation properties.

The rest of this paper is organized as follows: the proposed flexible load model and supply model are presented in Section II. 
The Proposed online algorithms and metrics for assessing the performance of the algorithms are presented in Section III. 
A case study is presented in Section IV, and conclusions are drawn in Section V.  

\section{Generation and Flexible Load Models} \label{sec:probform}

\subsection{Supply Model} \label{sec:supmod}
We consider two sources of supply for the dynamic matching market platform: 1) upstream grid supply (GS), and 2) distributed renewable energy sources (D-RES) in the distribution network. 
We assume that upstream grid supply, denoted by $p_t$, is sufficiently large and that it is priced at $c$ \$/unit of energy. 
The D-RES, such as PV solar, are by nature variable and uncertain, and their availability depends on weather, e.g., solar irradiance. 
Let us denote the D-RES generated at time $t$ by $S_t$, which is governed by a discrete-time stochastic process. 
We assume that the process $S_t$ is independent and identically distributed (i.i.d) and that it is known to the market platform. Denote the expectation with respect to all sources of randomness by $\mathbb{E}[.]$.
We denote the mean and standard deviation of D-RES $S_t$ by $ \mu_{s,t} =:\mathbb{E}[S_t], \sigma_{s,t} =: \mathbb{E}[(S_t - \mu_{s,t})^2]$, where $S_t$ is upper bounded by a constant $\overline{S}$. 

\subsection{Flexible Load Model} \label{sec:conmod}
Let us denote the number of loads which arrive at the platform at time $t$ by an independently and identically distributed stochastic process $n_t$, which is upper bounded by a constant $\overline{n}$. 
The mean and standard deviation of $n_t$ are respectively denoted by $\mu_{n,t} = \mathbb{E}[n_t]$, $\sigma_{n,t} = \mathbb{E}[(n_t - \mu_{n,t})^2$. 
Denote the set of loads that arrive at the platform by $\mathcal{K}$. Each load $k \in \mathcal{K}$ is characterized by three parameters $\{a^k,d^k,b^k\}$, where $a^k$ is the arrival time of the load, and $d^k$ is the specified deadline time to serve the load.
The parameter $b^k$ is the criticality of load $k$, which represents the rate at which a load's willingness to pay decreases over time. 
The heterogeneity of the loads lies in the differing deadlines and criticality. When load $k$ arrives in the platform it reports its service deadline $d^k$ and the value $b^k$. We note that loads of different types can arrive at the same time. This paper assumes that the loads report truthfully on arrival. The utility per unit of energy, shown by $\pi^k_t$, represents the load's willingness to pay for energy, and is defined as:
\begin{align} 
\pi^k_t  = c - b^k(t-a^k), ~~~ \pi^k_t > 0, \ \forall t, \ a^k \leq t \leq d^k
\label{eq:ass-wtp}
\end{align}

In \eqref{eq:ass-wtp}, the load's willingness to pay is less than or equal to the grid supply price $c$. 
This is reasonable considering that the grid supply is available at this price at all times. 
We assume without loss of generality that each load represents a unit of energy demand. 
This is because a load that exceeds a unit of energy can be treated as multiple loads of the same type.
From now on we drop the subscript $t$ in the moments of the random variables $S_t$ and $n_t$, since they are i.i.d. Also, we denote the combined standard deviation of the number of  renewable based generation and load arrivals by $\sigma = \sqrt{(\sigma_s)^2 + (\sigma_n)^2}$.


\section{Online Matching Algorithm}
This section proposes a pair of online algorithms to implement dynamic matching markets for two different generation-load scenarios in distribution systems. The objective of the proposed online algorithms is to maximize the social welfare of matching in the distribution system, subject to serving the loads in the market. 
The matching market for the distribution system operates for a duration of $T$ with time steps spaced equally at an interval $\Delta t$. The loads arrive in a sequential fashion governed by the stochastic process described in Section \ref{sec:conmod}. 
The generation from D-RES at any instant $t$ is given by the model described in Section \ref{sec:supmod}. At an instant $t$, the market maker can decide to match the energy demands of the loads that are currently active for the increment of time $\Delta t$ to D-RES or the grid supply or wait till later to match it. The market maker can make this decision only based on the information of the stochastic process that governs the future load arrivals and D-RES generation. 

Denote the energy matched to load $k$ at time $t$ by $q^k_t$, where $q^k_t \in \{0,1\}$, and the unit cost incurred by the platform for providing $q^k_t$ by $c^k_t$. The variable $q^k_t$ is the decision variable of the matching algorithm. 
We denote the energy utilized from the renewable supply at time $t$ by $s_t$, where $s_t \leq S_t$. 
Given these definitions, the social welfare for servicing the loads is defined as the sum of the utility of the loads minus the cost incurred by the market to serve the loads. The social welfare, $W$, is formulated as: $W := \sum_{k \in \mathcal{K}} (\pi^k_t - c^k_t)q^k_t$. 
Thus, the objective of the online algorithm is formally stated as follows:
\begin{equation}
 \max \ \mathbb{E}[W] \ \text{s.t.} \ \sum_k q^k_t = p_t + s_t \ \forall \ t.
\label{obj}
\end{equation}

We use $M_\sigma$ to denote an online algorithm for solving problem \eqref{obj}.
For a given realization (scenario) of load arrivals and D-RES generation for the full horizon, problem \eqref{obj} is solvable in polynomial time.
The so-called oracle optimal algorithm, denoted by $M_o$, is the optimal solution of the optimization problem for the complete information of load arrivals at each time step, their deadline and criticality and D-RES generation at each time step for the full horizon. Hence the oracle optimal algorithm achieves the maximum possible social welfare. 
We use the oracle algorithm as the benchmark for measuring $M_\sigma$'s relative performance, using the metric competitive ratio (CR) defined as follows. Denote the social welfare achieved by the platform's matching algorithm $M_\sigma$ over the horizon $T$ by $W[M_\sigma]$ and similarly denote the social welfare achieved by the oracle algorithm by $W[M_o]$. The CR for matching algorithm $M_\sigma$ is given by:
\begin{equation} \frac{\mathbb{E}[W[M_\sigma]]}{\mathbb{E}[W[M_o]]} \left( \text{Competitive Ratio (CR)} \right)
\end{equation}

We propose the following indicators based on the CR for measuring the effectiveness of an algorithm: (i) {\it convergence to optimality (CO)} as randomness reduces to zero, (ii) {\it deviation from optimality (DO)} measured as a function of combined standard deviation $\sigma$, which are formally defined below.

\begin{definition}Matching algorithm is said to achieve {\it Convergence to Optimality}, if the expected welfare  $\mathbb{E}[W[M_\sigma]]$ converges to $\mathbb{E}[W[M_o]]$ (i.e., CR converges to 1) as $\sigma \rightarrow 0$
\end{definition}

\begin{definition}{\it Deviation from Optimality} is the function $D(\sigma)$ such that: $\frac{\mathbb{E}[W[M_\sigma]]}{\mathbb{E}[W[M_o]]} \geq 1 - D(\sigma).$
\end{definition}

In particular we are interested in determining an upper bound to $D$ of the form $\sigma^r$. If $D \leq O(\sigma^r)$ then we say $r$ is the convergence rate. The notation $O(.)$ denotes that the term that accompanies the argument as a factor is a constant and does not scale with the problem's time horizon $T$. We say that the rate of deviation is linear if $r= 1$. 
We note that convergence is only a necessary property for being effective in managing the uncertainty in generation and loads. Deviation from optimality is a more well rounded measure as it describes the variation in the competitive ratio as the randomness varies.  


\subsection{Online Algorithm for the Case $\mu_n < \mu_{s}$} \label{sec:als}
We call the online algorithm we present for this case by $M_1$. 
This algorithm, presented in detail in Algorithm \ref{alg:M1}, approximately {\it matches the load with the highest criticality among the currently active loads to the available renewable supply. Any remaining load with an immediate deadline is matched to the grid supply.}

\begin{algorithm}[!th]
\floatname{algorithm}{}
\caption{\bf \strut Matching Algorithm $M_1$}
\label{alg:rh-Alg}
\setlength{\leftmargini}{0.1in}
\begin{enumerate}
\item At $t$, order the currently active loads ($m$ of them) such that $b^1 \geq b^2 \geq b^3 ... \geq b^m$, where if $b^{k-1} = b^{k}$ then $d^{k-1} < d^k$.
\item Match the $S_t$ units of D-RES to the first $S_t$ units of load in the above list. Call this matched set $\mathcal{I}_s$.
\item Match loads in the set $\mathcal{I}_{g} = \{i \vert \ i \notin \mathcal{I}_s, \ \exists j \in \mathcal{I}_s \ \text{s.t.} \ \pi^i_t >\pi^j_t \}$ to GS.
\item Match any remaining load with $d^k = t$ to GS.
\item t = t+1. GOTO 1.
\end{enumerate}
\label{alg:M1}
\end{algorithm}

Next we present Theorem \ref{thm:lb-cr-onpol1-arg1-g} that describes the properties of algorithm $M_1$. 

\begin{theorem}
{\it When $\mu_n < \mu_{s}$, the online algorithm $M_1$ satisfies CO and
\begin{equation}
\frac{\mathbb{E}[W[M_\sigma]]}{\mathbb{E}[W[M_o]]} \geq 1 - O\left(\sqrt{\sigma^2_n + \sigma^2_{s}}\right).
\end{equation}
}
\label{thm:lb-cr-onpol1-arg1-g}
\end{theorem}

Proof of Theorem 1 is provided in Appendix A. 
Algorithm $M_1$ is a ``greedy" algorithm as it tries to maximize the welfare it can gain at the current time by matching the most critical loads to the renewable supply generated at the current time. 
We note that the algorithm does not match any of the remaining loads, unless they have an immediate deadline, and they remain active. 
The online algorithm will achieve the optimal welfare if renewable supply is available to supply the waiting load, which may not be the case if adequate renewable supply is not available. 
Therefore, the social welfare attained by the algorithm can deviate from the optimal welfare that the oracle optimal achieves in certain instances. 
What we have shown is that the deviation from the oracle optimal is at least $O(\sigma)$. Hence, the rate at which the deviation varies, i.e., $r = 1$. This suggests that $r = 1$ is achievable when $\mu_n < \mu_{s}$. 

\vspace{-5pt}
\subsection{Online Algorithm for the Case, $\mu_n \geq \mu_{s}$} \label{sec:ags}
We start with a brief argument for why algorithm $M_1$ fails to satisfy CO for this case. Algorithm $M_1$ waits to serve a load until the renewable generation is available to supply the load. 
When $\mu_n > \mu_{s}$, the total amount of renewable energy generated over a large duration of time would fall short of the number of active loads during this period. 
Thus, in this case, algorithm $M_1$ would fail, with a high probability, to find renewable supply for certain loads. 
It is straightforward to show that this probability approaches to one as the randomness goes to zero. 
Consequently, algorithm $M_1$ can incur a net loss relative to the optimal welfare with probability one as the randomness goes to zero. In fact {\it when $\mu_n \geq \mu_{s}$, algorithm $M_1$ fails to satisfy CO and when customers which arrive later have higher criticality} $\text{CR} \geq 1 - O(\mu_n - \mu_{s}) - O\left(\sqrt{\sigma^2_n + \sigma^2_{s}}\right).$ 


Here we modify algorithm $M_1$, and develop Algorithm $M_2$ for the case when $\mu_n \geq \mu_{s}$. 
Algorithm $M_2$ is the following: {\it do the same steps as in $M_1$. In addition, match up to $\mu_n-\mu_{s}$ of the remaining loads that just arrived to the grid supply} (see Algorithm \ref{alg:M2}). This additional commitment on arrival ensures that the algorithm trivially satisfies CO. 
We present the properties of this algorithm as Theorem 2 below.

\begin{algorithm}[!th]
\floatname{algorithm}{}
\caption{\bf Matching Algorithm $M_2$}
\label{alg:rh-Alg}
\setlength{\leftmargini}{0.2in}
\begin{enumerate}
\item At $t$, order the currently active loads ($m$ of them) such that $b^1 \geq b^2 \geq b^3 ... \geq b^m$, where if $b^{k-1} = b^{k}$ then $d^{k-1} < d^k$. 
\item Match the $S_t$ of D-RES to the first $S_t$ units of load in the above list. Call this matched set $\mathcal{I}_s$.
\item Match loads in the set $\mathcal{I}_{g} = \{i\vert \ i \notin \mathcal{I}_s, \ \exists j \in \mathcal{I}_s \ \text{s.t.} \ \pi^i_t > \pi^j_t\}$ to GS.
\item Match up to $\mu_n-\mu_{s}$ of the remaining loads in the ordered list and of those that just arrived to the GS.
\item Match any remaining load with $d^k = t$ to GS.
\item t = t+1. GOTO 1.
\end{enumerate}
\label{alg:M2}
\end{algorithm}

\begin{theorem}
{\it When $\mu_n \geq \mu_{s}$, the online algorithm ($M_2$) satisfies CO and when customers which arrive later have higher criticality
\[ \frac{\mathbb{E}[W[M_\sigma]]}{\mathbb{E}[W[M_o]]} \geq 1 - O\left(\mu_n - \mu_{s}\right)-  O\left(\sqrt{\sigma^2_n + \sigma^2_{s}}\right).\]
}
\label{thm:lb-cr-onpol3-arg1-g}
\end{theorem}

Proof of Theorem 2 is provided in Appendix B. 
The main feature of algorithm $M_2$ is that it matches an additional set of loads that just arrived to the grid supply. The additional commitment on arrival ensures that the platform services certain loads earlier for which it could have failed to find renewable supply to service at a later time. This ensures that the algorithm satisfies convergence to optimality. From Theorem \ref{thm:lb-cr-onpol3-arg1-g} it follows that, under certain condition, the upper bound to the deviation from optimal welfare varies linearly with $\sigma$ but with an offset that is $O(\mu_n - \mu_{s})$. We note that the rate of deviation is for all practical purposes linear under these conditions when $\mu_n - \mu_{s}$ is very small. 

\section{Case Study}
Renewable generation profiles used in the simulation are derived from PV solar generation from 9:00 AM to 5:00 PM on the 10th and 19th of May -- each of these days are representative of different variation profiles observed during the month of the May -- in the state of California and then scaled down to a hypothetical generation capacity of $0.8$ MWh to arrive at the profile for the distribution system. 
The grid supply price is set to be the average retail price for residential customer in the U.S., which is $\$ 0.13/\text{kWh}$. 
The maximum limit of the number of loads that can arrive at a particular time is determined by the scenario. The number of loads at any instant $t$ is randomly drawn from the set of integers given by $1$ to the upper limit for the respective scenarios. The size of each load is set to be $0.1$ MWh and deadline and criticality of loads are arbitrarily assigned.

We consider the following standard baseline algorithms for comparison: (i) earliest deadline first (EDF), and (ii) matching renewable energy to loads with the maximum willingness to pay and any remaining load with immediate deadline to GS (MH). 
The simulations are run for a horizon of $T = 10$ for the following scenarios: (i) when the mean of D-RES is greater than the mean of the load arrivals and (ii) the scenario when this relation is reversed. For each of these scenarios we also consider two sub-scenarios: (i) where the variations are moderate and (ii) where the variations are large. 
Each scenario is repeated for $3000$ trials and the expected welfare is estimated by averaging across the trails. 
Table \ref{tab:welfare-alg} provides the values of the estimated expected welfare in the scenarios for the two benchmark algorithms EDF and MH, the oracle algorithm $M_o$ and the proposed algorithm $M_\sigma$ (without step (3)). 
The results in Table \ref{tab:welfare-alg} demonstrates that the proposed algorithm outperforms the benchmark algorithms in being the closest to the oracle algorithm in all the scenarios.

\begin{table}[h]
\centering
\caption{Expected Welfare of Algorithms}
\resizebox{0.45\textwidth}{!}{
\begin{tabular}{|c|c|c|c|c|} 
 \hline
Scenario & $\mathbb{E}[W[MH]]$ & $\mathbb{E}[W[EDF]]$ & $\mathbb{E}[W[M_o]]$ & $\mathbb{E}[W[M_\sigma]]$\\
 \hline
  $\mu_n < \mu_s$, small $\sigma$ & $882.2 \$$ & $886 \$$ & $893.5 \$$ & $892.9 \$$ \\ 
 \hline
  $\mu_n < \mu_s$, large $\sigma$ & $853.6 \$$ & $860.5 \$$ & $875.8 \$ $ & $872.8 \$$\\
 \hline
  $\mu_n > \mu_s$, small $\sigma$ & $965.6\$ $ & $941.4 \$ $ & $1000.6\$ $ & $993.4\$ $ \\
 \hline
  $\mu_n > \mu_s$, large $\sigma$ & $960\$ $ & $944.8\$ $ & $1005.3\$$ & $985.4\$ $ \\
 \hline
\end{tabular}}
\label{tab:welfare-alg}
\end{table}

\section{Conclusion}
In this paper, we designed online algorithms for dynamic matching markets in power distribution systems whose objective is to maximize social welfare. We proposed two indicators for characterizing the effectiveness of an online algorithm across scenarios (i) convergence to optimality (CO) as the randomness goes to zero and (ii) deviation from optimality (DO) measured as a function of the standard deviation, $\sigma$, of the distribution of renewable supply plus the number of loads that arrive on the platform. 
Under this notion of performance we presented a pair of algorithms that are effective for two distinct sets of generation-load scenarios. 
The contributions of this paper lies primarily in proving new theoretical results on the design and performance of online algorithms for dynamic matching markets in power distribution systems.


\bibliographystyle{IEEEtran}
\bibliography{Platform}

\section*{Appendix A: Proof of Theorem \ref{thm:lb-cr-onpol1-arg1-g}}
Let $W_{rs}$ be the welfare obtained from matching loads to D-RES and $W_{gs}$ be the welfare obtained from matching loads to GS. The proof entails the following steps.

{\it Step (i)}: We show that $W[M_\sigma] = W_{rs} + W_{gs} \geq W[M_o] + \overline{W}_{gs}$, where $\overline{W}_{gs}$ is the amount that the grid pays to the platform. Firstly, the welfare $W_{rs}$ cannot be increased by any sequence of feasible shifts of the matching of the renewable units, where the initial step in the sequence is an unmatching of a renewable unit, a shift to a load matched to GS is not allowed, and the final step could be matching of an unmatched renewable unit. Call this remark (O-1). This is because the loads that are more critical are matched to the renewable units before the loads that are less critical are matched to renewable units, which as a result of step ($3$) necessarily have a lesser willingness to pay (when compared to the willingness to pay at the time of matching of the more critical load) 
and so any such sequence of shifts will at the best result in the same value for $W_{rs}$.

Suppose a load is matched to GS at its deadline or before then the marginal welfare generated from this load is given by $(\pi_{d_m} - c)$, where $d_m$ is the time when the load is matched to GS. 
Consider all such loads that has been matched to the GS. Denote the set of such loads to be $\Theta_{gs}$. Then 
\begin{align}
 W[M_\sigma] & = W_{rs} + \sum_{\theta \in \Theta_{gs}} (\pi^\theta_{d_m} - c) \geq W[M_o] -  \sum_{\theta \in \Theta_{gs}} c \nonumber\\
 & =  W[M_o] + \overline{W}_{gs}.
\end{align}

The last inequality follows from the fact that $W_{rs} \geq W[M_o] - \sum_{\theta \in \Theta_{gs}} \pi^\theta_{d_m}$, which follows from (O-1) and the fact that any increase in $W_{rs}$ by shifting the matching of the loads from grid supply to renewable supply 
is less than $\sum \pi^\theta_{d_m}$, where the summation is over all the loads in $\Theta_{gs}$. The last observation follows from (i) that the loads that get matched to the grid supply at their deadline (step ($4$) of Algorithm $1$) are the less critical loads with properties similar to those in remark (O-1) and so the increase in $W_{rs}$ from shifting the matching of a renewable unit from a higher critical to such a load minus the decrease in welfare from shifting the matching of the higher critical loads will be lesser than $\pi_{d}$ at the deadline of the less critical load to which a renewable unit is shifted to; because the last higher critical load to be shifted in the sequence of shifts can only be shifted to a renewable unit after the deadline of the less critical load, 
and (ii) that the set $\mathcal{I}_g$ matched in step ($3$) of Algorithm $1$ includes the set of all the other less critical loads not included in point (i) and that could have been matched by the oracle optimal instead of the loads in the set $\mathcal{I}_s$ and so shifting a renewable unit to such a load will only add as much as $\pi_{d_m}$ to $W_{rs}$.



{\it Step (ii)}: We show that $\mathbb{E}[\overline{W}_{gs}] \geq -O(T\sqrt{\sigma^2_n + \sigma^2_{s}})$. 
We denote probability by $\mathbb{P}[.]$ and the indicator function by $\mathbb{I}[.]$. Let $\delta d_m$ be the minimum time interval after which a load that arrives at the platform can get matched to the grid supply. It follows from Algorithm 1 that a load which is active at time $t$ is matched to grid supply at time $t+\delta d_m$ only if 
\begin{equation}
\sum_{l = t}^{t+\delta d_m} S_l < \sum_{l = t}^{t+\delta d_m} n_l + \delta n,
\label{eq:condn-match-to-grid}
\end{equation}
where $n_l$ is the number of loads that arrive at time $l$, $\delta n$ are the number of loads that arrived before $t$ and are active at $t$, and $t+\delta d_m$ is the time the load active at $t$ is matched to the grid supply. 

Let $t_w \leq t$ be the last time instant less than or equal to $t$ when the loads that arrived before $t_w$ were not active at $t_w$. Then it follows that $\delta n \leq \sum_{l=t_w}^{t-1} n_l -  S_l$ and follows that the maximum number of loads which are active at $t$ and can be matched to grid supply at time $t+\delta d_m$ is upper bounded by 
\begin{equation}
(n_t -  S_t) + \delta n \leq \sum_{l=t_w}^{t} n_l -  S_l.
\label{eq:max-no-match-to-grid}
\end{equation}

We define the following two quantities:
{\small \begin{equation}
I_t = \mathbb{I}\left\{\sum_{l = t_w}^{\delta d_m + t} S_l < \sum_{l = t_w}^{\delta d_m + t} n_l\right\}, P_t = \mathbb{P}\left\{\sum_{l = t_w}^{\delta d_m + t} S_l < \sum_{l = t_w}^{\delta d_m + t} n_l\right\}, \nonumber
\end{equation}}

Then from Eq. \eqref{eq:condn-match-to-grid} and Eq. \eqref{eq:max-no-match-to-grid} it follows that the amount that the grid pays to platform at time $t' = \delta d_{m} + t$ for serving loads, $\overline{W}_{gs}(t')$, is lower bounded as given by
\begin{equation}
\overline{W}_{gs}(t') \geq c\left(\sum_{l=t_w}^t S_l -  n_l\right)\mathbb{I}\left\{\sum_{l=t_w}^t S_l - n_l < 0 \right\}I_t. \nonumber
\end{equation}
{\small\noindent This implies that
\begin{equation}
\mathbb{E}[\overline{W}_{gs}(t')\vert t_w] \geq c\mathbb{E}\left[\left(\sum_{l=t_w}^t S_l - n_l\right)\mathbb{I}\left\{\sum_{l=t_w}^t S_l - n_l < 0 \right\}I_t \right]. \nonumber
\end{equation}}
\noindent Since $(\mu_{s} - \mu_n) > 0$ in this case, we get that
\begin{align}
\mathbb{E}[\overline{W}_{gs}(t')\vert t_w] \geq & c\mathbb{E}\left[\left(\sum_{l=t_w}^t (S_l - \mu_{s}) - (n_l - \mu_n) \right) \times \right. \nonumber \\
& \left. \mathbb{I}\left\{\sum_{l=t_w}^t S_l - n_l < 0\right\}I_t\right]. \nonumber
\end{align}
\noindent Then using Cauchy Schwartz inequality we get that
\begin{align}
\mathbb{E}[& \overline{W}_{gs}(t')\vert t_w] \geq -c\sqrt{t-t_w+1}\left(\sqrt{\sigma^2_a + \sigma^2_{s}}\right) \times \nonumber \\
&\sqrt{\mathbb{P}\left\{\sum_{l=t_w}^t S_l - n_l < 0 \right\}}\sqrt{P_t}. \nonumber
\end{align}
\noindent The probability factor $P_t$ can be rewritten as follows:
\begin{align}
& P_t = \mathbb{P}\left\{\sum_{l = t_w}^{\delta d_m + t} S_l - \sum_{l = t_w}^{\delta d_m + t} n_l < 0\right\} \nonumber\\
& = \mathbb{P}\left\{\sum_{l = t_w}^{\delta d_m + t} \frac{(S_l - n_l - \mu_{s} + \mu_n)}{\left(\mu_n - \mu_{s}\right)} <  (\delta d_m+t-t_w+1)\right\}.\nonumber
\end{align}
\noindent Then using Hoeffding's inequality we get that
\begin{equation}
\sqrt{P_t} \leq \exp{\left\{-\frac{(\mu_{s} - \mu_n)^2(\delta d_m + 1)}{(\overline{n} + \overline{S})^2}\right\}} = e.
\label{eq:def-e}
\end{equation}
\noindent Using the fact that $\sqrt{t} \exp^{-at} < O(1)$ and applying Hoeffding's inequality to $\mathbb{P}\left\{\sum_{l=t_w}^t S_l - n_l < 0 \right\}$ we get that
\begin{equation}
\mathbb{E}[ \overline{W}_{gs}(t')\vert t_w]\geq -O\left(\sqrt{\sigma^2_n +\sigma^2_{s}}\right)\sqrt{P_t}.
\end{equation}
Then using Eq. \eqref{eq:def-e} we get that
\begin{equation}
\mathbb{E}[\overline{W}_{gs}(t')] \geq -O\left(\sqrt{\sigma^2_n + \sigma^2_{s}}\right)e. \nonumber
\end{equation}
\noindent Hence
\begin{equation}
\mathbb{E}[\overline{W}_{gs}] = \sum_{t = 1}^{T} \mathbb{E}[\overline{W}_{gs}(t)] \geq -TO\left(\sqrt{\sigma^2_n + \sigma^2_{s}}\right)e.
\end{equation}
This completes Step $2$. From the definition of CR it follows that
\begin{align}
\frac{\mathbb{E}[W[M_\sigma]]}{\mathbb{E}[W[M_o]]} & = \frac{\mathbb{E}[W_{rs}] + \mathbb{E}[W_{gs}]}{\mathbb{E}[W[M_o]]} \geq \frac{\mathbb{E}[W[M_o]] + \mathbb{E}[\overline{W}_{gs}]}{\mathbb{E}[W[M_o]]} \nonumber\\
& \geq 1 - \frac{TO\left(\sqrt{\sigma^2_n + \sigma^2_{s}}\right)e}{\mathbb{E}[W[M_o]]}.
\end{align}
\noindent Let us lower bound $\mathbb{E}[W[M_o]]$:
\begin{equation}
\mathbb{E}[W[M_o]] \geq \mathbb{E}\left[\sum_{t = 1}^{T} c \min\{ n_t , S_t \}\right] =  n_{sm} cT,
\end{equation}
\noindent where $\mathbb{E} \min\{ n_t , S_t \} = n_{sm}$. This implies that
\begin{equation}
\frac{\mathbb{E}[W[M_\sigma]]}{\mathbb{E}[W[M_o]]} \geq 1 - \frac{TO\left(\sqrt{\sigma^2_n + \sigma^2_{s}}\right)e}{n_{sm} c T} = 1 - \frac{O\left(\sqrt{\sigma^2_n + \sigma^2_{s}}\right)e}{n_{sm}} \nonumber
\end{equation} 
The property CO follows trivially from the lower bound derived above. $\blacksquare$

\section*{Appendix B: Proof of Theorem \ref{thm:lb-cr-onpol3-arg1-g}}
Let $W_{rs}$ be the welfare generated by algorithm 2 from matching loads to the renewable supply and $W_{gs}$ the welfare generated from matching loads to the GS. Similar to the steps in the proof of Theorem \ref{thm:lb-cr-onpol1-arg1-g} we get that 
\begin{align}
W[M_\sigma] & = W_{rs} + \sum_{\theta \in \Theta_{gs}} (\pi_d - c) \geq W[M_o] -  \sum_{\theta \in \Theta_{gs}} c \nonumber\\
 & = W[M_o] + \overline{W}_{gs}.
\end{align}

In this case $\overline{W}_{gs}$ can be divided in to two parts. One part corresponds to the payment made by the grid for the load that is matched on arrival as per step 4 of Algorithm $2$ to the GS, $\overline{W}_{gs1}$. The other part corresponds to the payment made by the grid for the other loads which are matched to GS, $\overline{W}_{gs2}$. It follows that \[ \mathbb{E}[W[M_\sigma]] \geq \mathbb{E}[W[M_o]]  + \mathbb{E}[\overline{W}_{gs1}(t)] + \mathbb{E}[\overline{W}_{gs2}(t)] \] Note that a load is matched on arrival to the grid supply by step $4$ of Algorithm $2$, at $t$, only when $S_t < n_t$, and up to $\mu_n - \mu_{s}$ are matched. This implies that \[\mathbb{E}[\overline{W}_{gs1}(t)] \geq -c\mathbb{E}[(\mu_n - \mu_{s})\mathbb{I}\{S_t < n_t\}].\] That is \[\mathbb{E}[\overline{W}_{gs1}(t)] \geq -c(\mu_n - \mu_{s})\mathbb{P}\{S_t < n_t\}.\]

\noindent {\it The lower bound for $\mathbb{E}[\overline{W}_{gs2}]$}: 
Let $\tilde{n}_t: n_t - \mu_n, \tilde{S}_t:= S_t-\mu_s$. If loads other than those matched by step $4$ of Algorithm $2$ and that arrive at $t$ are matched to the GS then it should be that $S_t < n_t - (\mu_n - \mu_{s})$, and only up to $\tilde{n}_t - \tilde{S}_t$ number of loads of the loads can be matched to the GS.
Thus, if any of the other loads do get matched to the GS then it is necessary that the cumulative sum of the renewable supply generated from its arrival time up to the matching time is insufficient to service the loads that arrive on the platform during this period, i.e., $\sum_{l=t}^{t+\delta d_m} S_l < \sum_{t}^{t+\delta d_m} n_l $, where $\delta d_m$ is the minimum time interval after arrival a load can get matched to the GS by a means other than the step $4$ of Algorithm $2$. The previous condition follows from the fact that the loads that arrive later have higher criticality and are ranked higher as per the ordering in Algorithm $2$.
Hence
\[\mathbb{E}[\overline{W}_{gs2}(t)] \geq c\mathbb{E}\left(\tilde{S}_t-\tilde{n}_t\right)\mathbb{I}\{\tilde{S}_t < \tilde{n}_t\}I_t,\]
where $\tilde{S}_t = S_t - \mu_s$ and $\tilde{n}_t = n_t - \mu_n$ and $t_w = t$ in $I_t$. Then from Cauchy Schwartz inequality we get that
\begin{equation}
\mathbb{E}[\overline{W}_{gs2}(t)] \geq -c\left(\sqrt{\sigma^2_s + \sigma^2_n}\right)\sqrt{\mathbb{P}\{\tilde{S}_t < \tilde{n}_t\}} \sqrt{P_t}.
\end{equation}

Combining the expression for the lower bound of $\mathbb{E}[\overline{W}_{gs1}]$ and $\mathbb{E}[\overline{W}_{gs2}]$ we get that
\begin{align} 
\mathbb{E}[W[M_\sigma]] \geq & \mathbb{E}[W[M_o]] -c\sum_{t = 1}^T(\mu_n - \mu_{s})\mathbb{P}\{S_t < n_t\} \nonumber \\
& - c\sum_{t = 1}^T\left(\sqrt{\sigma^2_{s} + \sigma^2_n}\right)\sqrt{P_t}.
\end{align}

Then following steps similar to the steps in the proof of Theorem \ref{thm:lb-cr-onpol1-arg1-g} we get that
\[ \frac{\mathbb{E}[W[M_\sigma]]}{\mathbb{E}[W[M_o]]} \geq 1 - \bar{c}_1\left(\mu_n - \mu_{s}\right)-  \bar{c}_2\left(\sqrt{\sigma^2_n + \sigma^2_{s}}\right),\]
where,
\[ \bar{c}_1 = \frac{\mathbb{P}\{S_t < n_t\}}{n_{sm}}, \bar{c}_2 = \frac{\sqrt{\mathbb{P}\{\tilde{S}_t  < \tilde{n}_t \}}\sqrt{P_t}}{n_{sm}}.\] CO follows from the definition of the algorithm. $\blacksquare$

\end{document}